# Single-spike superradiant free-electron lasers

**Authors:** Ke Feng[1✉], Runshu Hu[1,2], Yanjie Ge[1,2], Hai Jiang[1], Kangnan Jiang[1,3], Xizhuan Chen[1,2], Hao Wang[1], Gaojie Zeng[1], Zixiao Guo[1,2], Guiyao Wang[1,2], Jia Wang[1], Song Li[1], Jiayan Gui[1], Kegui Xia[1], Zongxin Zhang[1], Xiaojun Yang[1], Yi Xu[1], Yuxin Leng[1], Luyang Yu[4], Wei Zhang[4], Zhiqiang Jiang[4], Bo Liu[4], Chao Feng[4], Wentao Wang[1✉], Zhentang Zhao[4] and Ruxin Li[1,3✉]

**Affiliations:**

[1] State Key Laboratory of Ultra-intense laser Science and Technology, Shanghai Institute of Optics and Fine Mechanics (SIOM), Chinese Academy of Sciences (CAS), Shanghai 201800, China.

[2] Center of Materials Science and Optoelectronics Engineering, University of Chinese Academy of Sciences, Beijing 100049, China.

[3] School of Physical Science and Technology, ShanghaiTech University, Shanghai 201210, China.

[4] Shanghai Advanced Research Institute, Chinese Academy of Sciences, Shanghai 201210, China.

✉Correspondence author. E-mail: fengke@siom.ac.cn; wwt1980@siom.ac.cn; ruxinli@mail.siom.ac.cn.

## Abstract

**Free-electron lasers (FELs) deliver high-brilliance radiation from the infrared to X-rays, enabling atomic-scale studies on femtosecond-to-attosecond timescales[1-5]. Yet conventional facilities rely on large, costly radio-frequency accelerators. Laser wakefield accelerators (LWFAs), combining high accelerating gradients with ultra-high peak currents, offer a compact alternative and open up new prospects for advanced FEL operation[6-9]. Here, we demonstrate a tabletop superradiant FEL driven by single-spike, pre-bunched electron beams from a LWFA. By manipulating the energy chirp during acceleration and compressing the bunch to sub-wavelength duration, we obtain coherent radiation extending into the vacuum-ultraviolet (VUV) regime within a remarkably compact 3-metre beamline—from gas target to radiator. Under optimized conditions, the measured photon yield reaches $2.8\times10^{10}$ per pulse after transmission through a 500-μm slit, with a reproducibility of approximately 70%. This approach uniquely exploits the energy chirp and high-current characteristics inherent to LWFA beams without recourse to exponential gain, substantially minimizing device length while obviating the need for external synchronization and offering enhanced**

**tolerance to jitter—thereby establishing a practical route toward efficient, compact FELs suitable for applications in academic, industrial, and clinical settings.**

## Main Text

The advent of ultra-intense laser pulses based on chirped pulse amplification (CPA), coupled with advances in gain media, has revolutionized science[10,11]. High-intensity lasers exhibit remarkable capabilities, inspiring new frontiers in strong-field physics, attosecond physics, warm dense matter, and advanced laser-driven particle accelerators[12]. Since the development of infrared and visible lasers in the 1960s[13], the realization of an X-ray laser has remained a major scientific objective[14]. X-ray lasers, generating high-intensity coherent light pulses at angstrom wavelengths with femto- to attosecond durations, provide unprecedented atomic-scale spatiotemporal resolution across various scientific disciplines. This capability opens new windows into electronic, atomic, and molecular phenomena, offering transformative opportunities for physics, chemistry, materials science, and biology[14-17]. Three primary approaches have been developed for generating X-ray lasers, population inversion based on atomic inner-shell transitions[14,18], high-harmonic generation (HHG) in gaseous and solid media[19-21], and X-ray free-electron lasers (XFELs)[1,2]. Owing to their ability to produce radiation with brightness many orders of magnitude higher than other X-ray sources, XFELs have emerged as unique high-brilliance instruments for scientific research[16,17]. However, these facilities rely on electron beams from conventional radio-frequency (RF) accelerators, requiring the construction of large-scale facilities extending from several hundred meters to kilometers and incurring high costs[3,22-25]. Downsizing such facilities is therefore of profound significance for broadening their accessibility and fostering transformative technological applications.

With the ultra-high accelerating gradients reaching several hundred gigavolts per meter, laser wakefield accelerators (LWFAs) have attracted growing interest since they were first proposal by Tajima and Dowson[6]. By harnessing the immense charge-separation fields of the excited wake trailing behind the driving laser, LWFAs can accelerate electrons to GeV energies over distances of merely a few millimeters to centimeters—thousands of times shorter than those required in conventional RF accelerators[26]. LWFAs are regarded as highly promising drivers for compact FELs[8,9,27], and the realization of such compact LWFA-driven FELs has been identified as one of the major challenges in this decade, as addressed in the European Plasma Research Accelerator with eXcellence In Applications (EuPRAXIA)[28]. Motivated by the immense potential of these future compact facilities, remarkable progress has been achieved through innovative schemes, yielding substantial improvements in beam quality and stability[29-32]. Building on recent advances in beam quality enhancement, lasing has been demonstrated in LWFA-driven FELs operating in the self-amplified spontaneous emission (SASE) mode at wavelengths of 27 nm[33], 40 nm[34] and 420 nm[35], as well as in the seeded configuration at 274 nm[36]. However, advancing compact XFELs toward the saturation regime remains challenging due to persistent limitations in electron beam quality and stability.

Unlike conventional lasers that rely on atomic population inversion, high-gain FELs employ a fundamentally different gain mechanism. In an FEL, a relativistic electron beam undulates through a periodic magnetic array (undulator), interacting with its self-emitted radiation that co-propagates along the beam axis. Through this sustained interaction, the electrons become density-modulated at the resonant wavelength—a phenomenon known as microbunching—which drives the exponential amplification of the radiation[2]. Pre-bunching the electron beam at a longitudinal scale shorter than the target wavelength prior to its entry into the FEL interaction region enables the generation of fully coherent radiation[37], accelerates the gain process[38], and facilitates high-power and high energy extraction efficiency with a tapering-enhanced superradiant configuration[39]. This enhanced, pre-bunched emission is described as superradiant FEL, characterized by the quadratic dependence of radiation power on bunch charge[40-42]. By circumventing the limitations on minimum pulse duration and saturation inherent to high-gain FELs, superradiant FELs driven by pre-bunched electron beams offer an attractive route toward coherent sources with higher peak power and shorter pulse duration[43,44]. Furthermore, this scheme provides an alternative pathway for generating coherent radiation in wavelength regimes and operational conditions where conventional stimulated-emission sources are impractical. Superradiant free-electron lasers (FELs) were first observed at infrared wavelengths and later extended to the extreme ultraviolet (EUV) regime through harmonic cascaded configurations[38,45]. More recently, superradiant FELs have been demonstrated to produce terawatt-scale attosecond X-ray pulses, marking a significant milestone toward the development of ultra-compact, high-brightness coherent light sources[43,44].

Considering a pre-bunched electron beam wiggling in a short undulator (typically shorter than four gain lengths), the generated radiation power can be expressed as $P = F(N) P_e \pi^2 b_n^2 I_p K_0^2 N_u / \left(2\gamma I_A \left(1+K_0^2\right)\right)$, where $F(N) = 2\left[\arctan\left(1/(2N)\right) + N \ln\left(4N^2/\left(4N^2+1\right)\right)\right]/\pi$, $N = k\sigma^2/L_u$ is the diffraction parameter, $P_e[\mathrm{TW}] = E[\mathrm{GeV}] I_0[k\mathrm{A}]$ is the electron-beam power, $k$ is the wave number of the radiation, $\sigma$ is the averaged transverse size of the electron beam, $L_u$ is the total length of the radiator, $b_n$ is the $n$th local bunching factor, $I_p$ is the beam current, $K_0$ is the dimensionless undulator parameter, $N_u$ is the number of undulator periods and $\gamma$ is the Lorentz factor of the electron beam[46,47]. Leveraging the intrinsically high peak current of electron beams generated by LWFAs, advanced pre-bunching techniques—such as bunch compression or harmonic generation—can further elevate the peak current into the range of tens to hundreds of kiloamperes.

Consider a pre-bunched electron beam with a bunching factor of 10%, an energy of 0.5 GeV, a peak current of 30 kA, and an average transverse size of 100 μm. When such a beam passes through a compact radiator of only ten periods, the estimated radiation power reaches approximately 4.7 GW, which is comparable to, and in some cases even exceeds, the saturation power of a conventional high-gain FEL[2]. Such super-radiant FELs, driven by the ultra-high-current electron beams inherent to LWFAs, offer a compelling pathway toward intense coherent radiation without relying on exponential gain, thereby paving the way toward the realization of truly compact and widely accessible systems.

In this work, we present the first proof-of-principle demonstration of single-spike superradiant FEL lasing driven by an LWFA-accelerated electron beam. By precisely controlling the intrinsic energy chirp during acceleration and subsequently compressing the beam to sub-wavelength scales in a tunable dispersive undulator, we generate coherent radiation with wavelengths extending into the vacuum-ultraviolet (VUV) in a downstream ten-period radiator. The system delivers up to $2.8\times10^{10}$ photons per pulse (measured after a 500-μm spectrometer slit), with a reproducibility—defined as the fraction of successful shots—of approximately 70% under optimized conditions, all within a footprint of less than 3.0 m from gas target to radiator. Scanning the radiation intensity against the dispersive undulator gap provides an effective in situ diagnostic for the beam energy chirp. By consolidating complex phase-space manipulation and radiation generation within a laboratory-scale platform, this work establishes the practical feasibility of compact LWFA-driven FELs and offers a new paradigm for accessible photon facilities.

The experiments were performed using the self-developed 200 TW / 1Hz laser system[48], with the experimental layout schematically shown in Fig. 1a. An 800 nm, 28 fs laser pulse was focused by an off-axis parabolic mirror with an *f*-number of 32 into a supersonic gas jet target, with the corresponding vacuum spot size of 35.2 μm at full width at half maximum (FWHM) and an energy contraction of 59% at $1/e^2$ (see Methods for detail). The gas target consists of a pure helium nozzle and a perforated baffle designed to generate a shock front. The resulting energy chirp of the electron beam is tunable via the laser evolution and beam-loading dynamics (see Methods for detail). The accelerated electron beams were diagnosed using a high-resolution spectrometer positioned approximately 4.5 m downstream from the gas target, and detailed specifications of the spectrometer are provided in the Methods. The measured spectra of the accelerated electron beam from nearly 2000 consecutive shots are presented in Extended Data Fig. 1. The electron beams exhibit relatively large energy spreads due to the imposed energy chirp, with an average peak energy of 395.3 MeV and an average bunch charge of 40.4 pC. The corresponding root-mean-square (RMS) fluctuations of the peak energy remain within 4.5%. Start-to-end simulations[49] were conducted (see Extended Data Figs. 2–4), showing reasonable agreement with the experimental results.

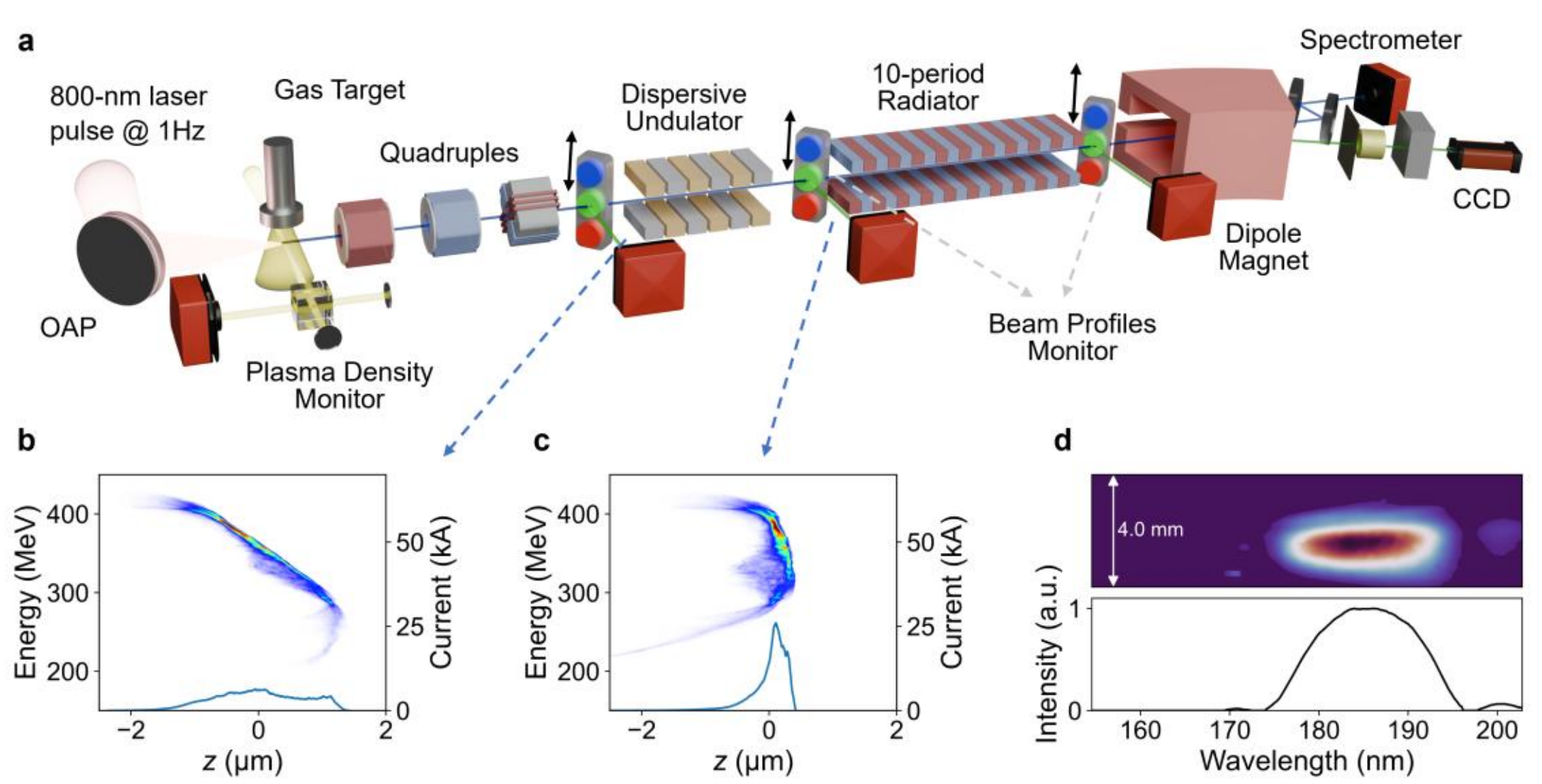

**Fig. 1 | Schematic layout of the single-spike superradiant FELs driven by LWFAs. a,** Layout of the experimental setup, with an overall length of 5.5 m from the gas target to the in-vacuum spectrometer. **b, c,** Simulated longitudinal phase-space distributions of the accelerated electron beam at the modulator entrance (**b**) and radiator entrance (**c**). **d,** Measured superradiant signal and its integrated spectrum, with a central wavelength of approximately 180 nm.

The accelerated electron beams are focused by a quadrupole triplet comprising two 5-cm-long permanent magnet quadrupoles with a magnetic gradient of 250 T/m and a 10-cm-long electromagnetic quadrupole with an adjustable gradient ranging from 0 to 80 T/m (see Methods for details). The beamline layout and the corresponding evolution of the electron beam parameters are presented in Extended Data Fig. 4. After focusing, the electron beams are directed into a five-period dispersive undulator with a period length of 6 cm and a variable gap. As the electrons wiggle through the undulator, lower-energy electrons exhibit larger oscillation amplitudes, resulting in longer path lengths that cause them to lag behind their higher-energy counterparts. By leveraging the positive energy chirp of the accelerated electron beam—wherein the bunch head possesses lower energy and can be readily tuned during the LWFA stage (see Extended Data Fig. 2)—the bunch duration can be effectively compressed through optimization of the dispersive undulator parameters. Consequently, the beam current evolves into a single-spike profile, attaining a peak current reaching several tens of kiloamperes with a bunch duration at the hundreds-of-attoseconds level, as illustrated in Figs. 1b and 1c. A typical shot of the obtained coherent radiation, centered at approximately 180 nm, is shown in Fig. 1d.

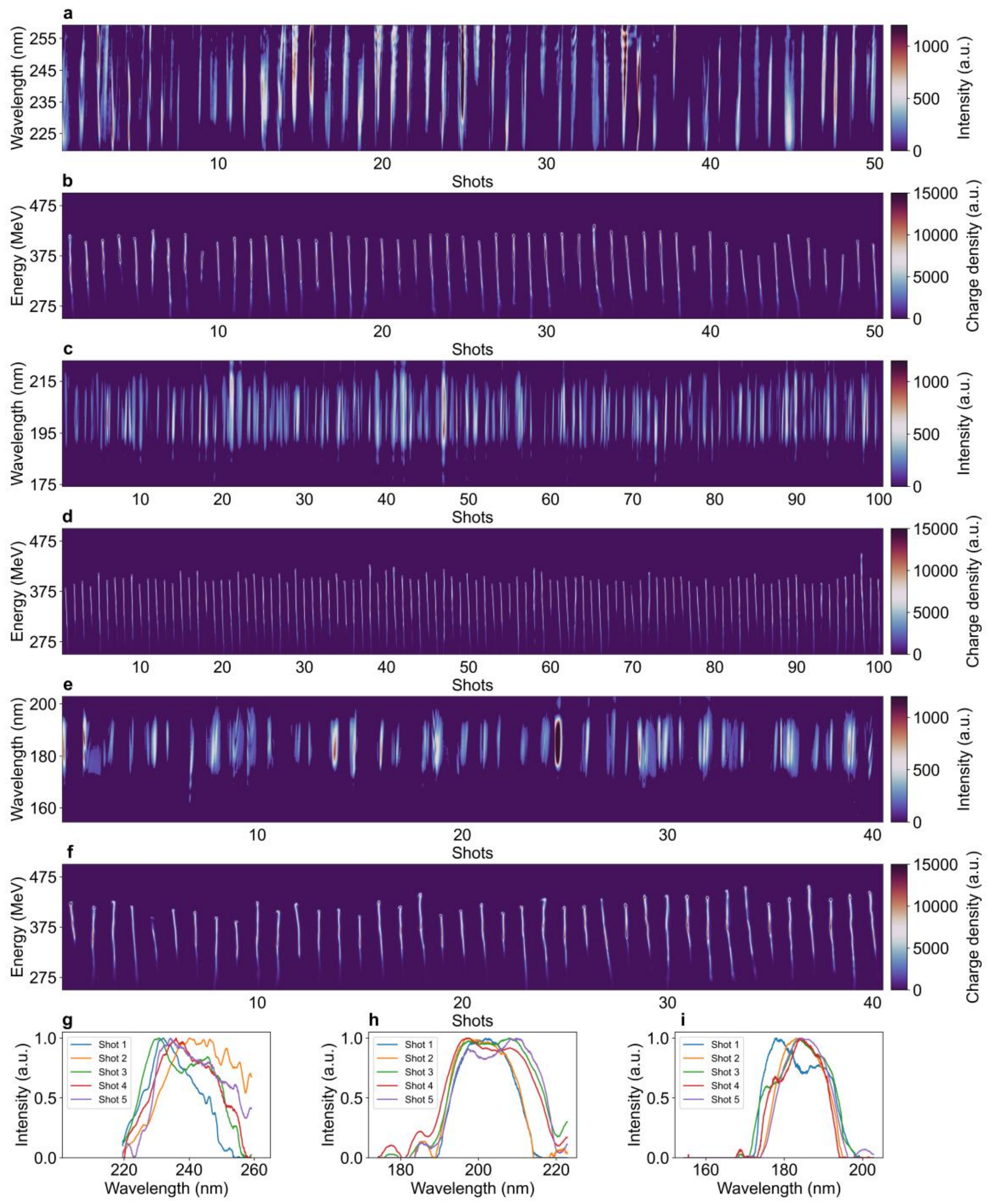

**Fig. 2 | Characteristics of superradiant emission and corresponding electron-beam spectra. a–f,** Experimentally measured superradiant spectra centered at 240 nm (**a**), 200 nm (**c**), and 180 nm (**e**), along with the corresponding electron-beam spectra (**b**, **d**, **f**). **g–i**, Integrated radiation spectra for five typical shots at 240 nm (**g**), 200 nm (**h**), and 180 nm (**i**).

This pre-bunched electron beam, with a bunch duration shorter than the radiation wavelength, was then directed into the downstream radiator for coherent emission. The radiator comprises ten periods, each 5 cm in length, with an adjustable gap that can be fine-tuned to match the electron beam energy to the target wavelength. For the experiments, the radiator gap was set to 21.0 mm (23.3 mm) for the target wavelength of 240 nm (180 nm), corresponding to undulator parameters of $K_{0R}$ = 2.6827 ($K_{0R}$ = 2.2622). The obtained radiation was reflected by a bandpass filter mirror to block the residual drive laser before being directed into an in-vacuum spectrometer through a 500-μm slit (see Methods for details). Figures 2a–2f present the measured superradiant emission across multiple shots, with central wavelengths of approximately 240 nm, 200 nm, and 180 nm, alongside the corresponding electron beam spectra. Figures 2g–2i show the integrated radiation spectra corresponding to five typical shots from Figs. 2a, 2c, and 2e. Notably, the observed spectra exhibit an ultrabroad bandwidth, a characteristic of the present superradiance scheme driven by an ultra-high-current electron beam in a few-period undulator, where the spectral width is primarily governed by the Fourier-transform limit of the ultrashort coherent pulse. Averaged over the spectra shown in Figs. 2g–2i, the bandwidth is 10.2%, which is in reasonable agreement with the theoretical prediction of $1/N_u$~10%, in line with the superradiant signature of the radiation. All radiation spectra presented here have been background-subtracted. For several shots at central wavelengths of 200 nm and 180 nm, weak sideband structures are discernible on both the longer- and shorter-wavelength sides of the main peak. These sidebands are mainly attributable to the wavelength-dependent reflectivity of the mirror.

Figures 3a and 3b illustrate the charge of the accelerated electron beam versus the total signal counts of the observed radiation at central wavelengths of 200 nm and 240 nm, respectively, with the corresponding statistical properties also provided. The maximum measured photon number reaches $2.8\times10^{10}$ per shot, obtained from background-subtracted CCD counts. The upper envelope of the data represents the maximum radiation energy for a given electron-beam charge, indicating the optimal compression and matching conditions. This envelope exhibits a distinctly nonlinear growth, a hallmark of the superradiant signature of the obtained radiation. Owing to shot-to-shot fluctuations in the matching conditions, however, the majority of shots fall below this envelope. Notably, in certain shots, no signal was observed despite sufficient charge, suggesting a mismatch in experimental conditions that led to incoherent spontaneous emission rather than coherent superradiance. As shown in Extended Data Fig. 5 (black dots, logarithmic scale), this incoherent emission is orders of magnitude weaker than its superradiant counterpart, falling below the detection background and thus unmeasurable under the current experimental conditions.

To identify the optimal conditions for bunch compression, the radiation intensity was systematically scanned as a function of the dispersive undulator gap, with the results presented in Fig. 3c. The green and red solid curves denote the average of the ten highest-count shots under each condition, selected from several tens of measurements at central wavelengths of 180 nm and 240 nm, respectively. The shaded regions correspond to the RMS fluctuations among these selected shots. The radiation intensity reaches its maximum at a dispersive undulator gap of 26 mm for both target wavelengths (180 nm and 240 nm), corresponding to a dispersive undulator parameter of $K_0$ = 2.8675. The total dispersive strength of the undulator and drift section is estimated to be approximately −7.2 μm, indicative of optimal pulse compression. This yields an inferred energy chirp of about 49.1 MeV/μm for the LWFA-accelerated electron beam (see Methods for detail). At gap settings below this value, the radiation intensity falls off markedly due to overcompensation of the energy chirp. Extended Data Fig. 6 presents the radiation spectra for 20 consecutive shots at both target wavelengths (180 nm and 240 nm), acquired under different dispersive undulator gaps. The reproducibility of the radiation exhibits a strong dependence on the dispersive gap, indicating that optimization of the bunch compression—and consequently entry into the superradiant regime—can be achieved by tuning the undulator's dispersive strength. All count data in Fig. 3c have been background-subtracted. Fluctuations observed under mismatched conditions are predominantly attributable to background noise (Extended Data Fig. 5).

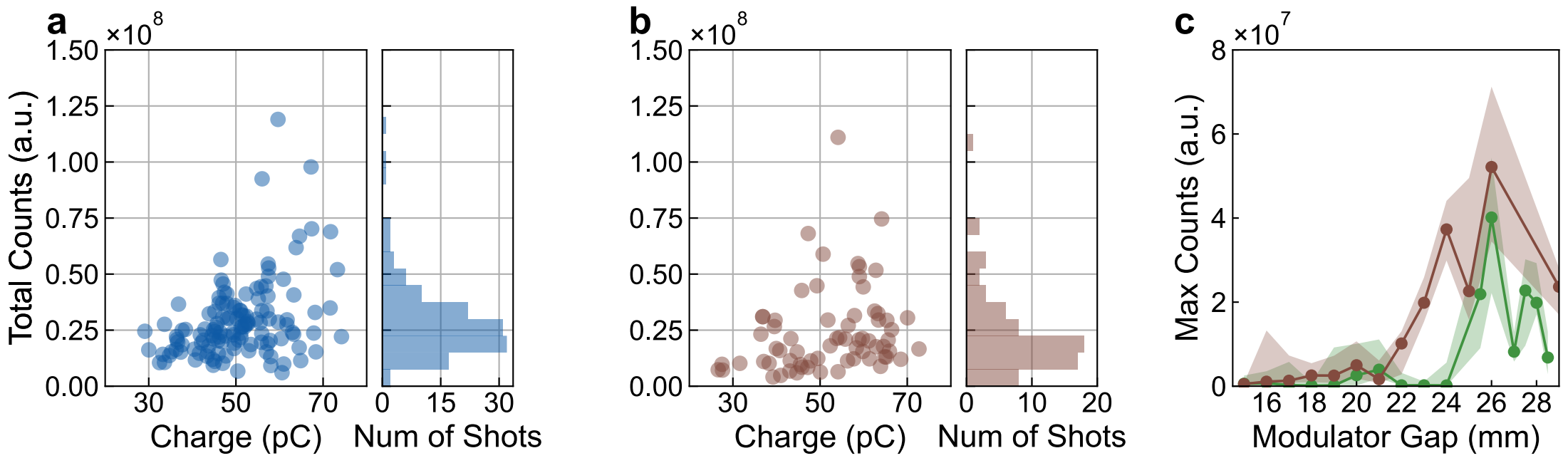

**Fig. 3 | Statistical properties of the superradiant emission. a, b,** Electron beam charge versus total signal counts measured by the CCD at the end of the beamline, together with the corresponding count histograms, for central wavelengths of 200 nm (**a**) and 240 nm (**b**). **c**, Total signal counts as a function of the dispersive undulator gap for central wavelengths of 180 nm (green solid line) and 240 nm (red solid line). Each data point represents the average of the ten highest-count shots, with the shaded areas indicating the range of the RMS values within these selected shots. The radiator gap was set to 23.3 mm for 180 nm and 21.0 mm for 240 nm, corresponding to undulator parameters of $K_{0R}$ = 2.2622 and $K_{0R}$ = 2.6827, respectively.

In conclusion, we have demonstrated a proof-of-principle single-spike superradiant FEL driven by a pre-bunched electron beam from an LWFA. By precisely controlling the intrinsic energy chirp of the LWFA-accelerated electron beam and achieving sub-wavelength bunch compression in a dispersive undulator, we generated coherent radiation at wavelengths down to 180 nm. At optimized settings, the maximum photon yield reaches $2.8\times10^{10}$ per shot after a 500-μm spectrometer slit, with a shot-to-shot reproducibility of approximately 70%. Crucially, this superradiant paradigm circumvents the fundamental bottlenecks that have long constrained compact coherent light sources. It fundamentally bypasses the stringent constraints on beam energy spread, emittance, and pointing stability inherent to high-gain SASE FELs, while simultaneously eliminating the need for an external seed laser and its associated demanding spatiotemporal synchronization. By fully exploiting the intrinsic energy chirp and ultra-high peak currents characteristic of the LWFA beams, our entire beamline—from gas target to radiator—is condensed to a footprint of less than 3.0 m while sustaining high output performance. Moreover, scanning the dispersive undulator parameter provides a highly effective in situ diagnostic for the beam energy chirp. As indicated in Extended Data Fig. 7, this scheme is extendable to the extreme ultraviolet, and no fundamental physical barrier prevents extending this chirp-managed, pre-bunched scheme to the X-ray regimes[50]. Our results thus establish a robust and scalable platform, offering a viable pathway toward next-generation compact coherent X-ray sources. Critically, by bringing such capabilities within reach of university laboratories and research centers, this approach could thereby accelerate scientific discovery and innovation across diverse fields—from materials science and chemical dynamics to structural biology.

## Methods

### Laser system

The experiments were performed using a self-developed 200 TW/1 Hz laser system based on chirped-pulse amplification (CPA)[48]. The system delivers pulses centered at 800 nm with a duration of 28 fs (full width at half maximum, FWHM), exhibiting an energy fluctuation of 0.55% (root mean square, RMS) and a pointing jitter of 1.5 μrad (RMS), as measured over 5400 consecutive shots during a 90-minute period[48]. The laser pulses were focused onto the gas target by an off-axis parabolic mirror with an *f*-number of 32, producing a vacuum focal spot of 35.2 μm (FWHM) and an energy contraction of 59% at $1/e^2$. In the presented experiments, the on-target pulse energy was approximately 3.43 J, corresponding to a normalized vector potential of $a_0 = 1.43$.

**Laser wakefield acceleration**

Laser wakefield acceleration was implemented utilizing a density-tailored plasma scheme based on shock-front injection. The supersonic gas target comprised a pure helium gas nozzle with a bore diameter of 4.5 mm, combined with a perforated baffle designed to shape the gas flow. This baffle consisted of a 2.5-mm-thick plate featuring a central orifice for laser passing through, with its distance from the nozzle adjustable to control both the shock-front position and, consequently, the effective acceleration length. To characterize the plasma density profile, a probe beam was directed perpendicularly through the gas target, and a Michelson-type interferometer equipped with a 4*f* optical imaging system was employed. The resulting plasma density profile are presented in Extended Data Fig. 2.

**Electron-beam diagnostics**

The accelerated electron beam was characterized using a magnetic spectrometer positioned 4.5 m downstream of the gas target. The spectrometer comprised a 60-cm-long tunable dipole magnet capable of delivering fields up to 1.5 T, coupled to a Lanex phosphor screen optically imaged onto a charge-coupled device (CCD, QImaging Retiga R3). For an electron beam with 0.1-mrad divergence, the spectrometer achieved an energy resolution of approximately 0.1% at 400 MeV. The angular acceptance was constrained to ±1.43 mrad, determined by the 8-mm minimum inner aperture of the downstream undulators, resulting in an energy measurement uncertainty of approximately 2.5% at 400 MeV. Absolute charge calibration was performed using BAS-TR imaging plates (IPs), with a measured uncertainty of 11%. The measured spectra of the accelerated electron beam from nearly 2000 consecutive shots are presented in Extended Data Fig. 1.

**Beamline components**

The beamline comprised a quadruple-triplet configuration, consisting two 5-cm-long permanent-magnet quadrupoles (PMQs), each providing a fixed gradient of 250 T/m and a 10-cm-long electromagnetic quadrupole (EMQ) with a variable gradient tunable from 0 to 80 T/m. The longitudinal spacing between the first PMQ and the gas target, as well as the inter-PMQ separation, was adjustable to optimize beam matching conditions. Both PMQs were mounted within the vacuum chamber on three-axis motorized translation stages, enabling precise alignment of their magnetic axes. To monitor the position and transverse profile of the accelerated electron beam, several beam profile monitors incorporating cerium-doped yttrium aluminum garnet (YAG:Ce) screens are installed along the beamline. The entire beamline spans approximately 3 m (5.5 m) from the gas target to the radiator (spectrometer), with all components coaxially aligned to the driving laser within a tolerance of ±100 μm. The layout of the beamline is presented in Fig.1 and Extended Data Fig. 4.

**Dispersive undulator and radiator**

The dispersive undulator is a planar undulator located 1.14 m downstream from the gas target, consisting of 5 periods with a period length of $\lambda_{mu}$ = 60 mm. Its magnetic gap was mechanically tunable from 11 to 30 mm, yielding a peak vertical field of $B_{m0}$ = 1.38 T at the minimum gap setting. The radiator, positioned 2.14 m downstream of the target, incorporated 10 periods of 50-mm length ($\lambda_{ru}$ = 50 mm) with an adjustable gap range of 10 to 30 mm, delivering a maximum horizontal field of $B_{m0}$ = 1.42 T at the 10-mm gap. Both undulator sections were interconnected by evacuated beam pipes with elliptical cross-sections, maintaining a semi-minor axis of 4 mm throughout the modulator and radiator interiors.

**Photon beam diagnostics**

The radiation signal was characterized using an in-vacuum spectrometer (Andor 234/302 monochromator) and an X-ray charge-coupled device (CCD, Andor Newton DO940P), both installed at the end of the beamline. Prior to detection, the radiation was reflected by a bandpass reflecting mirror to shield against residual main laser, and then sent to the spectrometer with the slit width of 500 μm. Spectral analysis was performed using a reflective grating with a groove density of 1200 lines per mm, providing a highest resolution of 0.1 nm at 200 nm. The grating exhibited diffraction efficiencies of 21.4% at 240 nm, 26.2% at 200 nm and 28.8% at 180 nm. Following dispersion, the spectrally resolved signal was imaged onto the X-ray CCD, which features a 2048× 512 pixel array with a pixel size of 13.5×13.5 μm. The corresponding quantum efficiency of the CCD is 17.8% at 240 nm, 19.8% at 200 nm and 20.4% at 180 nm.

**Particle in cell (PIC) simulation**

Numerical simulations were conducted using the spectral, quasi-three-dimensional particle-in-cell (PIC) code FBPIC with a co-moving simulation window. The simulation parameters were matched to the experimental conditions[51,52]. The simulation domain spanned 80 μm in the longitudinal direction and 160 μm in the transverse direction, with a longitudinal grid size of 40 nm, a transverse grid size of 80 nm and 16 macroparticles per cell. The driving laser was modelled as a linearly polarized Gaussian pulse with central wavelength of $\lambda_L$ = 800 nm, a normalized vector potential of $a_0$ = 1.43, a focal waist radius of $\omega_0$ = 29.9 μm, and a pulse duration of 28.0 *f*s (FWHM). The target medium was assumed to be pre-ionized, with the density profile reconstructed from experimental diagnostic results, as shown in Extended Data Fig. 2.

**Macroparticle up-sampling and particle tracking**

To accurately model the FEL interaction, a sufficient number of macroparticles must be ensured. A joint cumulative distribution function approach[53] was therefore employed to up-sample the initial 12.7 k macroparticles from the FBPIC simulation to 4.0 million macroparticles. The phase-space distributions of the electron beam prior to and following up-sampling are presented in Extended Data Fig. 3, together with the corresponding current profiles, energy spreads, and wavelength-dependent bunching factors. Notably, the initial bunching present in the raw FBPIC output was entirely eliminated through the up-sampling procedure. This suppression of spurious pre-bunching precludes unphysical coherent spontaneous emission (CSE) in subsequent FEL simulations. Subsequent particle tracking throughout the entire beamline was performed using the Ocelot code[54], incorporating space-charge and second-order transport effects. The initial beam distribution for these tracking simulations was derived from the up-sampling output, with results shown in Extended Data Fig. 4.

**Estimation of electron beam energy chirp**

The energy chirp of the accelerated electron beam can be estimated from the combined dispersion coefficients of the vacuum drift section and the dispersive undulator. As shown in Fig. 3, the radiation is optimized at a dispersive undulator gap of 26 mm (corresponding to a dispersive undulator parameter of $K_0$ = 2.8675), indicating that optimal pulse compression has been achieved. The dispersion coefficients contributed by the dispersive undulator and the drift section were calculated as $R_{56}^{\mathrm{U}} = L_u / \gamma_0^2 \left(1 + K_0^2 / 2\right)$ and $R_{56}^{\mathrm{D}} = L_D / \gamma_0^2$, respectively. Based on the present beamline parameters, the total dispersion coefficient is estimated to be −7.2 μm. The corresponding energy chirp of the electron beam is thus determined to be 49.1 MeV/μm, which is in reasonable agreement with the simulation results of 43.3 MeV/μm presented in Extended Data Figs. 2 and 3.

**FEL simulation**

FEL interaction dynamics were simulated using the three-dimensional (3D) full-wave code Mithra[55], operating within a Lorentz-boosted framework. The input electron beams were obtained from the FBPIC and Ocelot codes, as previously described. For the optimized configuration targeting at the most efficient bunching (see Extended Data Fig. 7), radiation was simulated at a target wavelength of 70 nm. The simulation domain spanned 4.6 mm × 4.6 mm × 24.2 μm in the horizontal, vertical, and longitudinal directions, with grid sizes of 20 μm, 20 μm, and 8 nm, respectively. The radiator featured a period length of 40 mm and an undulator parameter of 1.35. Notably, this undulator parameter differs slightly from that used in the experiments presented herein, as the simulation focused on the capability of the proposed scheme to generate shorter-wavelength radiation. As shown in Extended Data Fig. 7, under the optimized operating conditions, the bunching factor reaches 4.5% at 70 nm, and the corresponding peak current increased to approximately 68 kA. According to start-to-end simulation, coherent radiation with a peak power of about 500 MW can be achieved at 70 nm in such a compact device. Further increases in the electron beam energy and charge are expected to yield higher-power radiation at even shorter wavelengths[50].

**Acknowledgements** This work was supported by the National Natural Science Foundation of China (Grant No. 12388102), the Strategic Priority Research Program of the Chinese Academy of Sciences (Grant No. XDB0890200), the National Natural Science Foundation of China (Grant Nos. 12225411 and 12474349), the National Key R&D Program of China (Grant No. 2025YFF0515000), the CAS Project for Young Scientists in Basic Research (Grant No. YSBR060), the CAS Youth Innovation Promotion Association (No. 2022242), Zhangjiang Laboratory, and the New Cornerstone Science Foundation through the XPLORER PRIZE.

**Author contributions** K. F., W. W. and R. L. conceived the project and planned the experiments. K. F. designed and managed the experiments. K. F., R. H., H. J., K. J., X. C., H. W. and G. Z. performed the experiments and collected the data. J. G., K. X., Z. Zhang., X. Y., Y. X. and Y. L. constructed and ran the Ti:sapphire laser system. L. Y., W. Z., Z. J., are responsible for equipment installation and debugging. K. F., R. H., Y. G., H. J. and G. W. analysed the experimental data. K. F., Y. G. and J. W. conducted the simulations. K. F., W. W.

and R. L. co-wrote the paper. The results here jointly discussed by K. F., R. H., Y. G, H. J., K. J., Z. G., S. L., B. L., C. F., W. W., Z. Zhao and R. L.

**Competing interests** The author declare no competing interests.

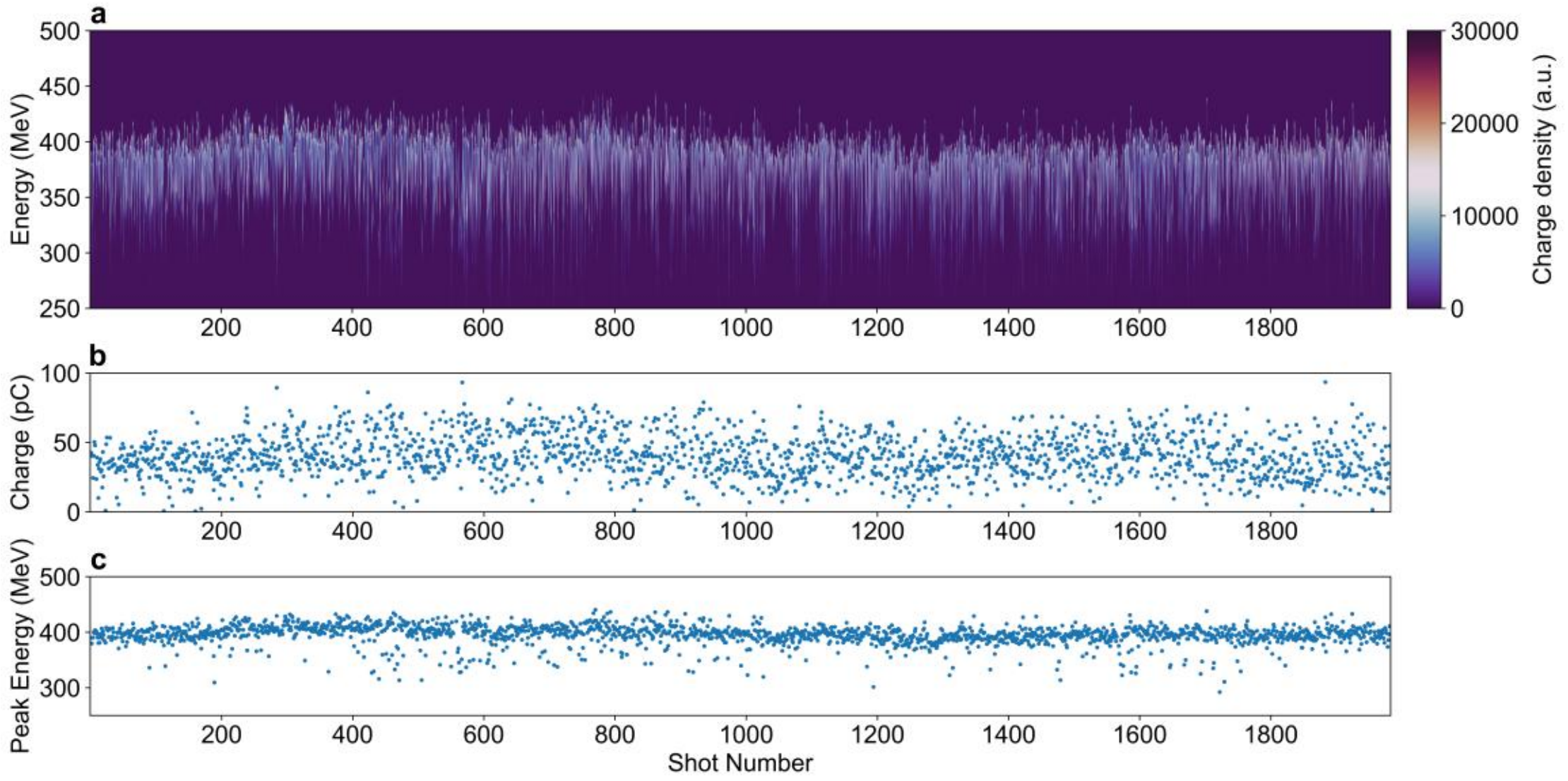

**Extended Data Fig. 1 | Electron beam properties measured at the end of the beamline. a**, Energy spectra of the accelerated electron beams measured over the course of a single day by the spectrometer located 4.5 m downstream from the gas target. **b, c,** Corresponding peak energy (**b**) and bunch charge (**c**) for the spectra shown in **a**. The accelerated electron beams exhibit an average peak energy of 395.3 MeV and an average bunch charge of 40.4 pC, with root-mean-square (RMS) fluctuations of 4.5% and 36.8%, respectively.

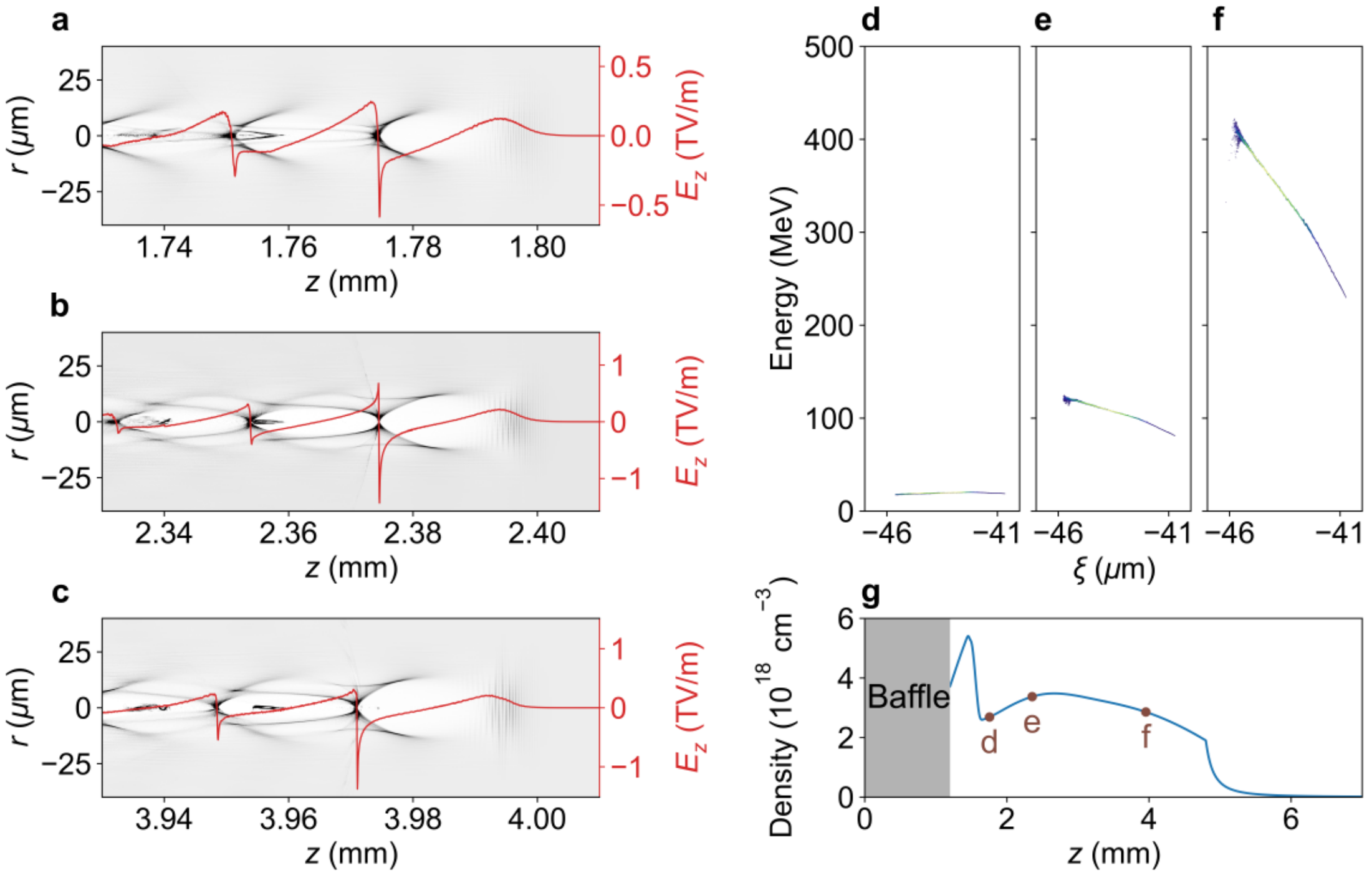

**Extended Data Fig. 2 | Particle-in-cell (PIC) simulation results for LWFA. a–c,** Snapshots of the electron density map and the on-axis accelerating field (red solid lines) at different longitudinal positions. **d–f,** Corresponding longitudinal phase spaces of the accelerated electron beam for the positions shown in **a–c**. **g,** Measured on-axis plasma density profile. The gray shaded area indicates the position of the baffle, and the markers denote the locations corresponding to **d–f**.

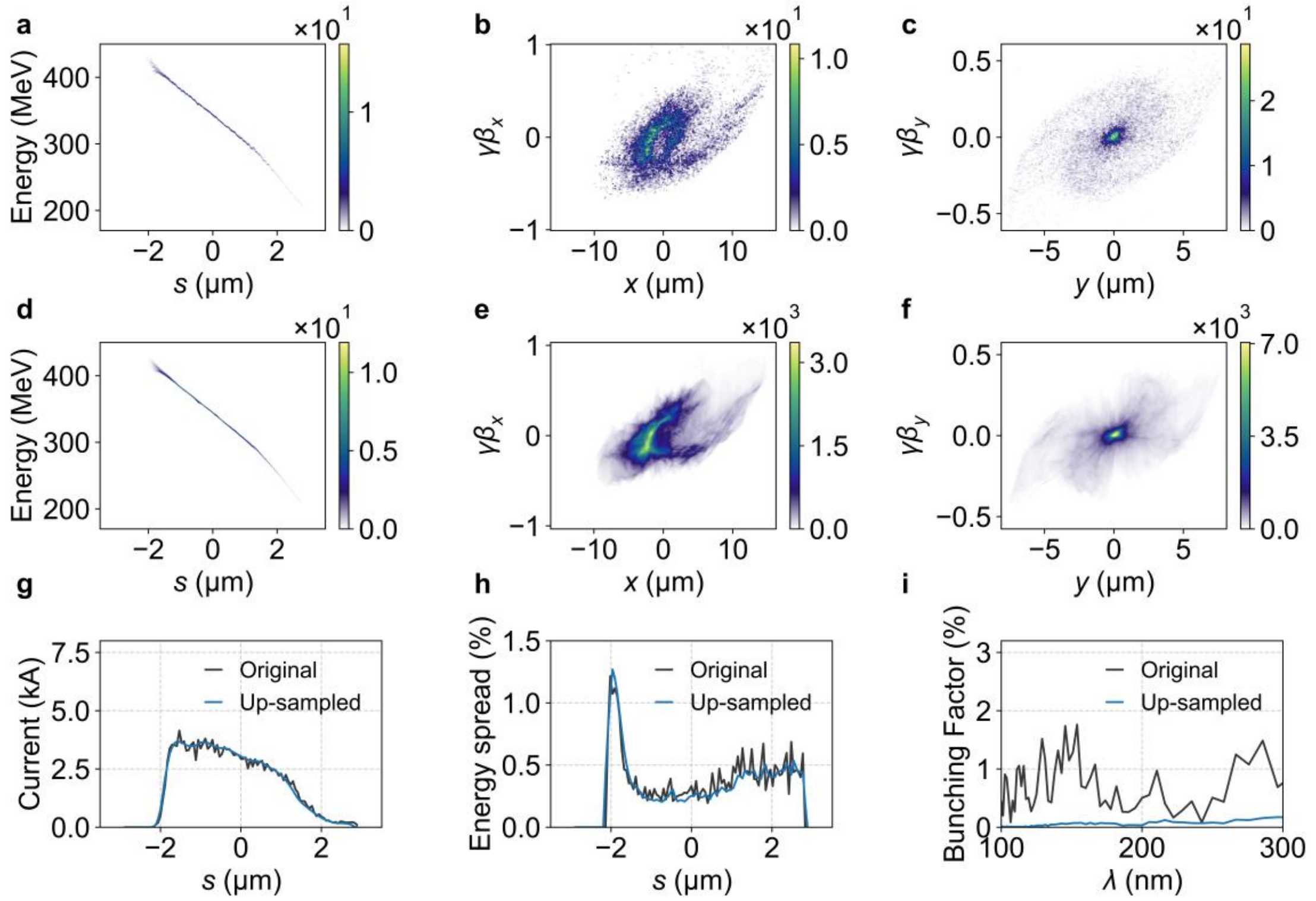

**Extended Data Fig. 3 | Simulated electron beam properties before and after up-sampling. a–f**, Longitudinal and transverse phase-space distributions of the electron beam as obtained from the FBPIC simulation (**a–c**) and after up-sampling (**d–f**). **g–i**, Comparison of the original (black solid lines) and up-sampled (blue solid lines) electron beam: current profile (**g**), slice energy spread along the bunch (**h**) and bunching factor as a function of wavelength (**i**). The number of macroparticles was increased from an initial 12.7 k to 4.0 million using a joint cumulative distribution function method.

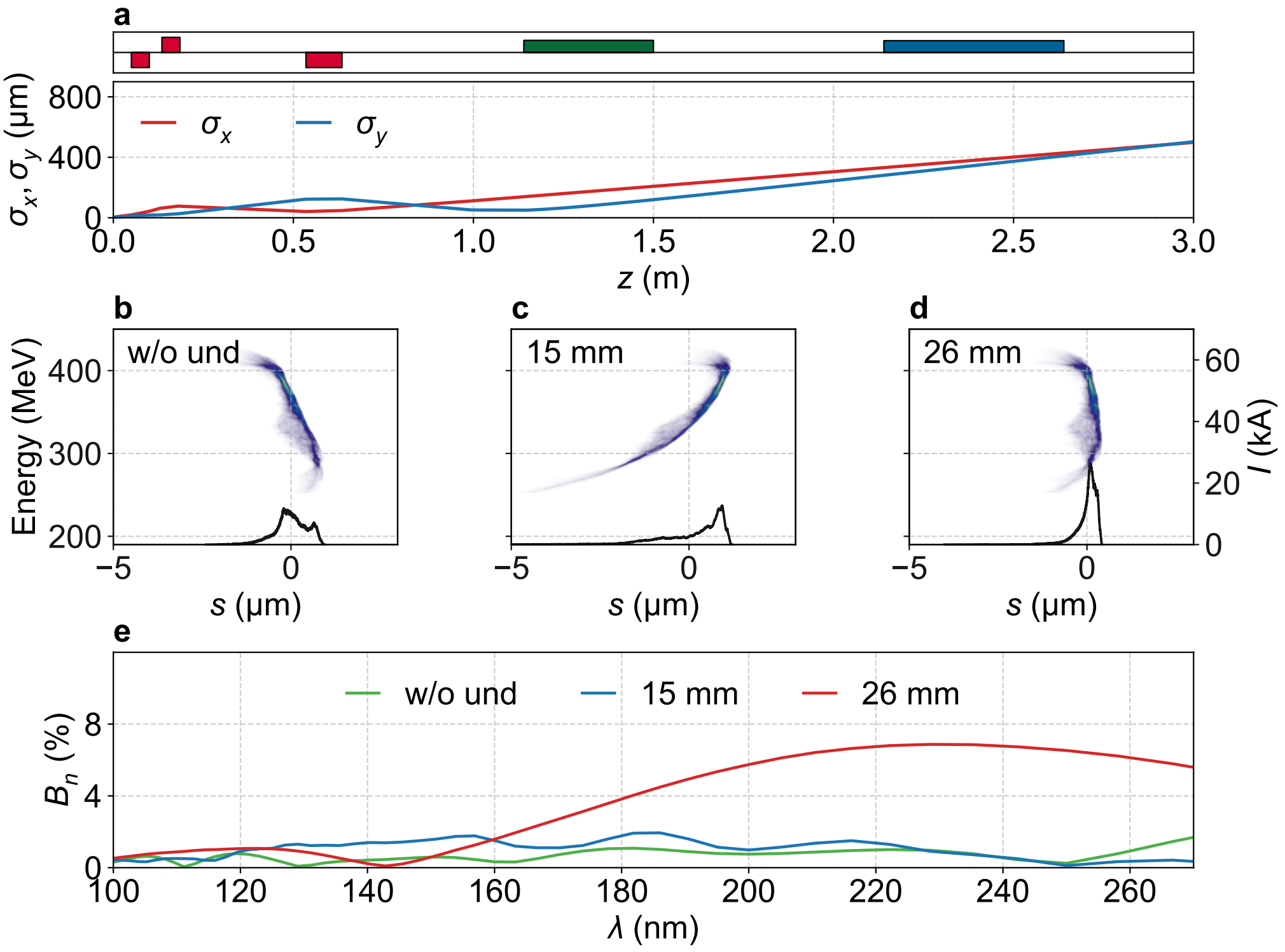

**Extended Data Fig. 4 | Properties of the accelerated electron beam throughout the beamline.** **a**, Beamline layout and evolution of the transverse beam size along the beamline. **b–d**, Longitudinal phase-space distributions and corresponding current profiles at the radiator entrance without dispersive undulator (**b**) and for dispersive undulator gaps of 15 mm (**c**) and 26 mm (**d**), respectively. **e**, Bunching factor as a function of wavelength for the different dispersive undulator gaps shown in **b–d**.

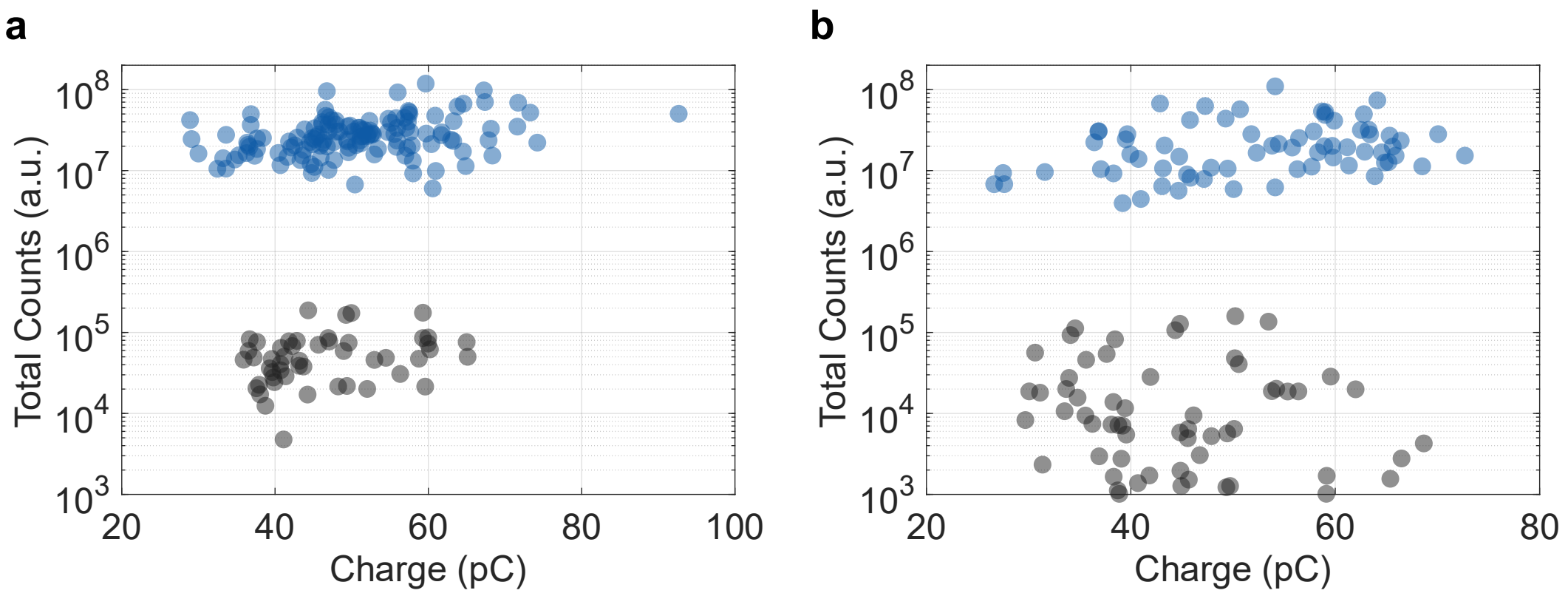

**Extended Data Fig. 5 | Dependence of radiation counts on electron beam charge. a, b,** Total measured counts after background subtraction for the super-radiance signal (optimal compression, blue dots, blue dots) and background (unmatched dispersion, black dots) as a function of beam charge at radiation wavelengths of 200 nm (a) and 240 nm (b), respectively.

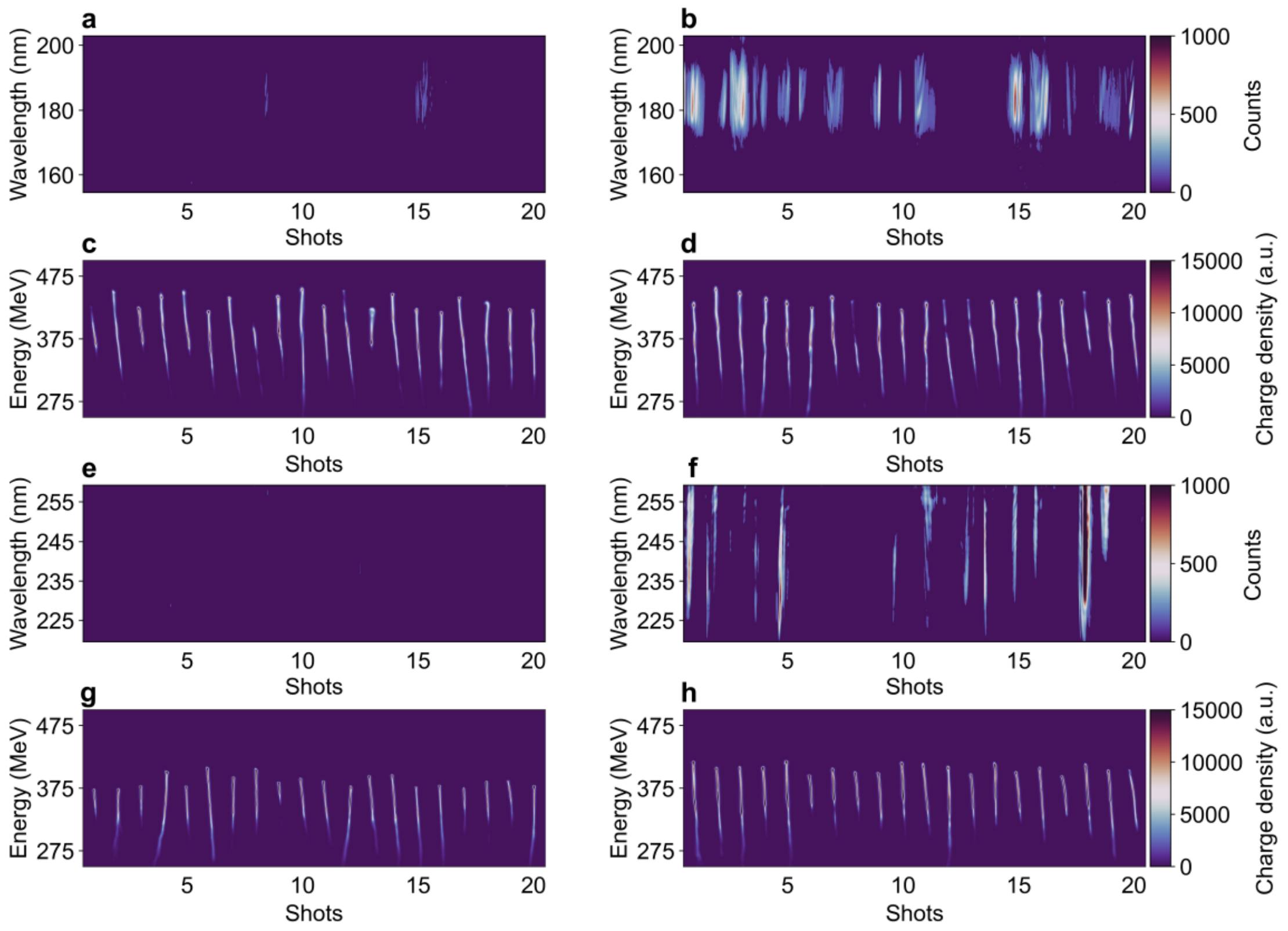

**Extended Data Fig. 6 | Dependence of superradiance on bunch compression conditions. a–d,** Radiation spectra (**a, b**) and corresponding electron beam spectra (**c, d**) for 20 consecutive shots measured at a radiator gap of 23.3 mm (~180 nm central wavelength), with dispersive undulator gaps of 15 mm (**a, c**) and 26 mm (**b, d**). **e–h,** Radiation spectra (**e, f**) and corresponding electron beam spectra (**g, h**) for 20 consecutive shots measured at a radiator gap of 21.0 mm (~240 nm central wavelength), with dispersive undulator gaps of 15 mm (**e, g**) and 24 mm (**f, h**). The reproducibility of superradiant emission exhibits strong dependence on the dispersive undulator gap, indicating that optimal bunch compression—and thus access to the superradiant regime—is achieved by tuning the dispersive strength.

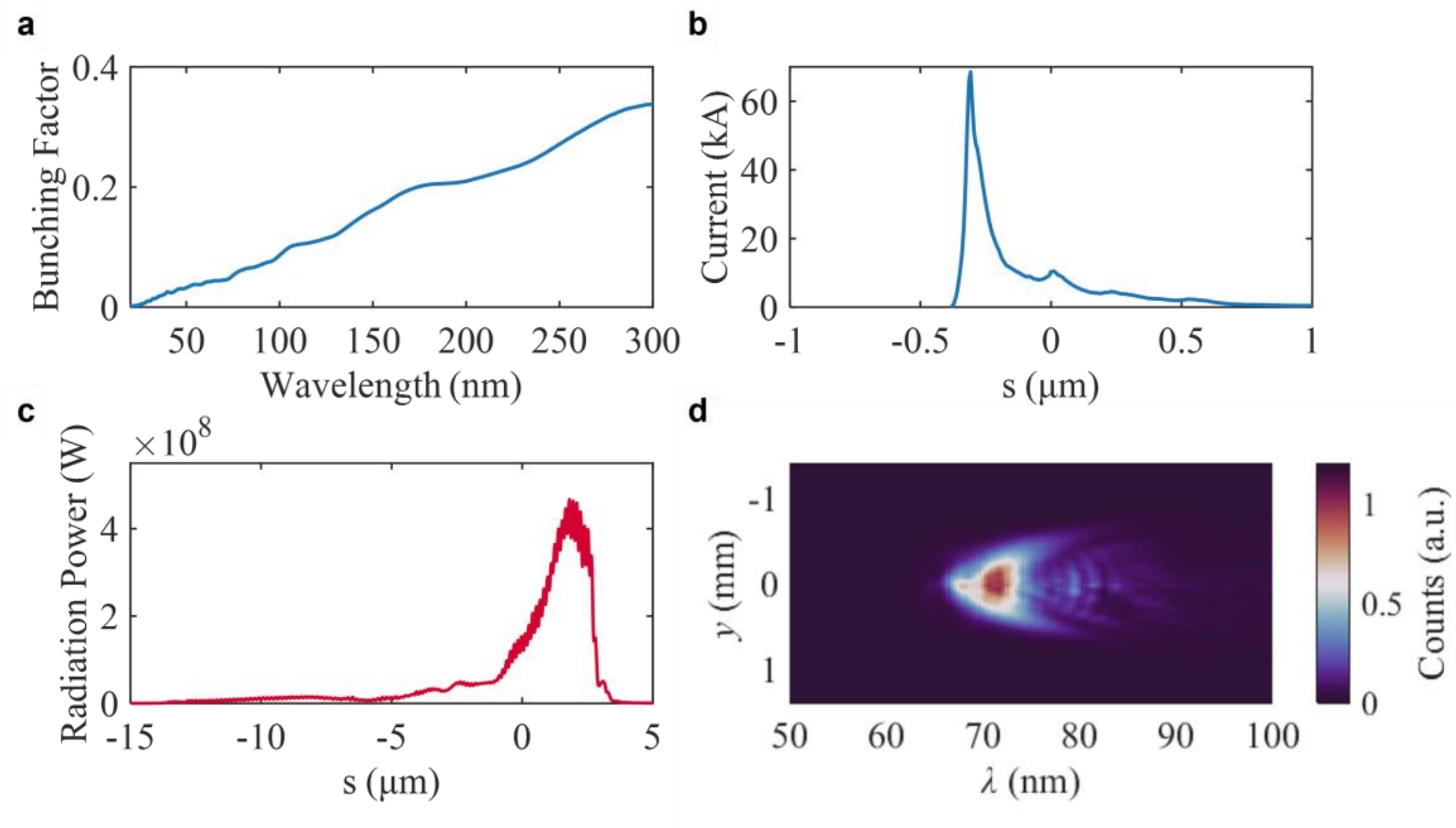

**Extended Data Fig. 7 | Start-to-end simulation results for coherent radiation with central wavelength of 70 nm. a–b,** The bunching factor as a function of the target wavelength (**a**) and the corresponding current profile of the electron beam (**b**) under the optimized beamline configuration targeting at the most efficient bunching. **c–d,** Simulated radiation power and the corresponding spectrum of the single-spike coherent radiation with central wavelength of approximately 70 nm.